\documentclass[10pt,leqno]{amsart}
\usepackage{graphicx}
\usepackage{indentfirst,csquotes}
\usepackage{amssymb,amsthm,amsmath}
\begin{document}
\newcommand{\eq}{\begin{equation}}                              
\newcommand{\eqe}{\end{equation}}             
\title{Even and odd self-similar solutions of the diffusion equation for infinite horizon } 
\author[L. M\'aty\'as]{ L. M\'aty\'as}
\author[I.F. Barna]{ I.F. Barna}
\address{Department of Bioengineering, Faculty of Economics, Socio-Human Sciences
and Engineering, Sapientia Hungarian University of Transylvania, Libert\u{a}tii sq. 1, 530104 Miercurea Ciuc, Romania, }
\email{matyaslaszlo@uni.sapientia.ro }
\address{ Wigner Research Center for Physics, 
Konkoly-Thege Mikl\'os \'ut 29 - 33, 1121 Budapest, Hungary }
\email{barna.imre@wigner.hu}

\maketitle


\begin{abstract}
 
 In the description of transport phenomena an important aspect represents the diffusion. In certain cases the diffusion may appear together with convection. In this paper we study the diffusion equation with the self similar
 Ansatz. With an appropriate change of variables we found original solutions of diffusion equation for infinite horizon. Here we present the even solutions of diffusion equation for the boundary conditions presented. 
For completeness the odd solutions are also mentioned as well, as part of the previous works. 
Finally, the diffusion equation with constant source term is discussed, which also has even and odd solutions, too.

\end{abstract} 

  \section{Introduction} 

 It is an evidence that mass diffusion or heat conduction is a fundamental physical process which attracted enormous intellectual interest for mathematicians, 
physicists and engineers in the last two century. The existing literature about mass and heat diffusion is immense, we just mention 
some basic textbooks \cite{crank, ghez, ben, lien, neu}. 

Regular diffusion is the corner stone of many scientific discipline like, surface growth \cite{KPZ,BaMa-KPZ1,BaSt1995}, reactions diffusion \cite{MaGa2005}
or even flow problems in porous media. In our last two papers we gave an exhaustive summary about such 
processes with numerous relevant reviews \cite{laci, imre1}.     

In connection with thermal diffusion \cite{Cannon1984,CoBe2011} it is also possible the presence of heat and mass transfer simultaneously, which may lead to cross effects \cite{MaTeVo2000}. 
Relevant applications related to general issues of heat transfer or engineering one may find in \cite{Thamby2011}.    
Important diffusive phenomena occur in universe \cite{MiAlRi2015} which is another field of interest.

The study of population dynamics or biological processes \cite{Murray,Al2023,Perthame2015} also involves diffusive processes, 
especially in spatial extended systems. 
In environmental sciences, the effects of spreading, distribution and adsorption of particulate matter or pollutants is also relevant \cite{SzeNeKe2017,GiSe2013,Pernyeszi2009,NeNe2009}. 
Furthermore from practical purposes diffusion coefficients have been measured in food sciences as well 
\cite{LvReCh2018}. 

As new applications in the last decades diffusion gained ground in social sciences as well. As examples we can mention diffusion of innovations \cite{innov1,innov2},  diffusion of technologies and social behavior \cite{tech} or even diffusion of cultures, humans or ideas  \cite{diff_gen1,diff_gen2}. 
Aspects related to diffusion one may also find in the theory of pricing \cite{Maz2018,Ede2014}. 
The structure of the network has also a crucial role which influences the spread of innovations, ideas or even computer viruses \cite{Reka-Barabasi-2002}. 
Parallel to such diffusion activities generalization of heat-transport equations were done by V\'an and coauthors \cite{van1} e.g. forth order partial differential equations (PDE)s were formulated to elaborate the problem of non-regular heat conduction phenomena.  
  Such spirit of the times clearly show that investigation of diffusion (and heat conduction) is still an important task. 

Having in mind that diffusion can be a general, three dimensional process beyond Cartesian symmetry here we investigate the one dimensional diffusion equation. The change in time of variable $C(x,t)$ is influenced by the presence of it in the neighbors: 
\begin{eqnarray}
\frac{\partial C(x,t)}{\partial t  } = D \frac{\partial^2 C(x,t)}{\partial x^2  }, 
\label{pde} 
\end{eqnarray}
where $D$ is the diffusion coefficient which should have positive real value. One assumes that $C(x,t)$ is a sufficiently smooth function together with existing derivatives, regarding both variables.  
  
In this general form one may observe that if $C(x,t)$ is a solution, then 
$C(x,t)+C_0$ is also solution, where $C_0$ is a constant. 

For finite horizon or interval, in case the concentration is fixed at the two ends $C(x=0,t)=C_0$ and $C(x=L,t)=C_0$ the solutions are 
\begin{equation}
C_k (x,t) = C_0 + e^{-D \frac{\pi^2 k^2 t}{L^2}} \cdot 
\sin\left( \frac{k \pi}{L} x \right), 
\label{sol_sin}
\end{equation}
where $k=1,2,3...$, it can be any positive integer number.  
In general, beyond $C_0$ any linear combination of the 
product of the exponent and sine for different $k$ is a solution.  
For finite horizon, in the case when the density is fixed to zero on both ends  
the solutions are changed to 
\begin{equation}
C_k (x,t) = C_0 + e^{-D \frac{\pi^2 n^2 t}{L^2}} \cdot 
\cos\left( \frac{n \pi}{L} x \right), 
\label{sol_cos}
\end{equation}
where $n=1,2,3...$, can be any positive integer number.  
Thanks to the Fourier theorem, with the help of Eq. (\ref{sol_sin}) and Eq. (\ref{sol_cos}) arbitrary diffusion 
profile can be approximated on a closed interval. 
These are well-known analytic results and can be found in any usual physics textbooks like 
\cite{crank,ghez}. 

In the present study - with the help of the self-similar Ansatz - we are going to present generic symmetric solutions 
for infinite horizon. These solutions have their roots at the very beginning of the theory, in the form of the Gaussian \cite{crank,ghez}:
\begin{equation} 
C(x,t) = \text{Const}. \cdot \frac{1}{\sqrt{t}} e^{-\frac{x^2}{4Dt}}.  
\label{classic}
\end{equation} 

For infinite horizon there are also certain works which present 
a given aspect of the diffusion, and it may arrive to 
a slightly more general aspect than the classical solution presented above 
\cite{bluman}.   

In the following we will go much beyond that point and will present and analyze completely new type of solutions. Finally, 
the corresponding Green's function will be given which makes it possible to handle physically relevant arbitrary initial condition problems.

 \section{Theory and Results}
In case of infinite horizon, when we want to derive the corresponding solutions we make the following self-similar transformation: 
\begin{eqnarray}
C(x,t) = t^{-\alpha} f\left(\frac{x}{t^{\beta}} \right)  = t^{-\alpha} f(\eta).   
\label{ansatz}
\end{eqnarray}
Note, that the spatial coordinate $x$ now runs along the whole real axis. \\ 
This kind of Ansatz have been applied by  
Sedov \cite{sedov} later also used by Raizer and Zel'dowich \cite{zeld} 
For certain systems Barenblatt applied it successfully \cite{barenb} as well.  
We have also used it for linear or non-linear partial differential equation (PDE) systems, which are from fluid mechanics \cite{imre2,BaPoLo2017,BaPoBa2022} or quantum mechanical systems \cite{imre3}. In certain cases the equation of state of the fluid also plays a role 
\cite{Csanad2010,BaMa2013}. 
Diffusion related applications of the self-similar analysis method can be found in relatively recent works, too \cite{NaSi19,Sa20,KaSuZh20}.    

The transformation takes into account the (\ref{classic}) formula, and 
before the function $f$, instead of $1/\sqrt{t}$ there is a generalized function $1/t^\alpha$, and in the argument of $f$, the fraction 
$x/t^\beta$ is possible, with a $\beta$ which should be determined later. 

We evaluate the first and second derivative of relation 
(\ref{ansatz}), and insert it in the equation of diffusion (\ref{pde}). 
This yields the following  ordinary differential equation (ODE) 
\eq
-\alpha t^{-\alpha -1} f(\eta)  - \beta   t^{-\alpha -1} \eta  
\frac{d f(\eta)}{d\eta}  =  D  t^{-\alpha -2\beta}  \frac{d^2f(\eta)}{ d\eta^2}.   
\label{ode}
\eqe

The reasoning is self-consistent if all three terms has the same 
decay in time. This is possible if  
\eq 
\alpha =     \textrm{arbitrary real number},   \hspace*{3mm} \beta = 1/2,
\eqe
and yields the following ODE   
\eq
-\alpha f - \frac{1}{2}\eta f' = D f''.  
\label{ode1}
\eqe
This ODE is a kind of characteristic equation, 
with the above presented change of variable. 
\begin{figure}  
\scalebox{0.55}{
\rotatebox{0}{\includegraphics{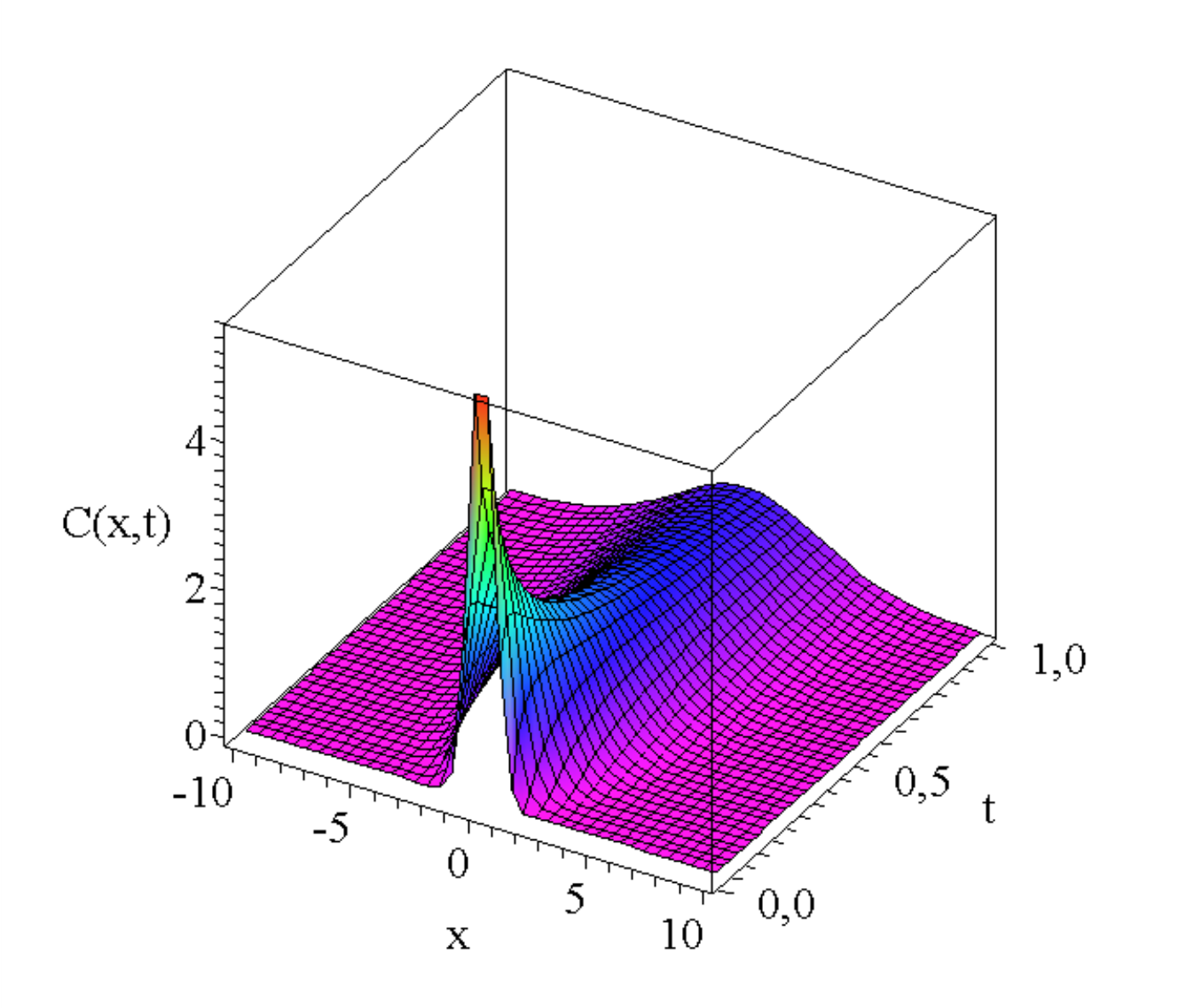}}}
\caption{The solution $C(x,t)$ for $\alpha = 1/2$ which is the usual Gaussian solution of Eq.  (\ref{egyketted}).}
\label{egyes}      
\end{figure}
One can observe that for $\alpha=1/2$ this equation can be written as 
\begin{equation}
- \frac{1}{2} \left( \eta f \right)'' = D f''.
\end{equation}     
If this equation is integrated once  
\begin{equation} 
\text{Const}_0 - \frac{1}{2} \eta f = D f',
\end{equation} 
where $\text{Const}_0$ is an arbitrary constant, which may depend on certain conditions related to the problem. If we take this $\text{Const}_0 = 0$, then one arrives to the generic solution 
\begin{equation} 
f = f_0 e^{-\frac{\eta^2}{4D}},
\end{equation} 
where $f_0$ is a constant. 
Inserting this form of $f$ in form of $C(x,t)$ given by Eq. (\ref{ansatz}) - for $\alpha=1/2$ as it was mentioned earlier - one gets an even solution for the space variable: 
\begin{equation} 
C(x,t) = f_0 \frac{1}{t^\frac{1}{2}} e^{-\frac{x^2}{4Dt}}.
\label{egyketted}
\end{equation}  
By this we have recovered the generic Gaussian solution, which can be seen on figure (\ref{egyes}). 

If we want to find further solutions, the equation (\ref{ode1}) has to be solved for general $\alpha$. 
The general solution for infinite horizon of (\ref{ode1}) can be written as: 
\begin{equation}
f(\eta) = \eta  \cdot e^{-\frac{ \eta^2}{4D} }  \left(  c_1 M\left[1-\alpha  , \frac{3}{2} , \frac{ \eta^2}{4D} \right]  + c_2 U\left[1 -\alpha , \frac{3}{2} , \frac{ \eta^2}{4D} \right]    \right),
\label{f_eta2}
\end{equation}
where $c_1$ and $c_2$ are arbitrary real integration constants and 
$M(,,)$ and $U(,,)$ are the Kummer's functions. For exhaustive details 
consult \cite{NIST}.  

If $\alpha$ are positive integer numbers, then both special functions $M$ and $U$ are finite polynomials in terms of the third argument 
$\frac{ \eta^2}{4D} $ 
\begin{equation} 
f(\eta) = \eta  \cdot e^{-\frac{ \eta^2}{4D} } 
\left( \kappa_0 +\kappa_1  \frac{\eta^2}{4D}  + ... + \kappa_{n-1} \cdot \left[\frac{\eta^2}{4D}\right]^{n-1}   \right). 
\end{equation}   
These gives the { \it{odd solutions} } of the diffusion equation for $\alpha=n$, (where $n$ positive integer), in terms of the space variable. 
It follows for the complete solution $C(x,t)$  
\begin{equation} 
C(x,t) = \frac{1}{t^n} f(\eta) =  
\frac{1}{t^n} \frac{x}{\sqrt{t}}  e^{-\frac{x^2}{4D t}} \cdot  
\left( \kappa_0 +\kappa_1  \frac{x^2}{4Dt}  + ... + \kappa_{n-1} \cdot \left[\frac{x^2}{4Dt}\right]^{n-1}   \right). 
\end{equation}   
These odd solutions have been studied thoroughly by M\'aty\'as and Barna in previous works (\cite{laci,imre1}) and for completeness, we present these solution in Appendix A. 

\begin{figure}  
\scalebox{0.65}{
\rotatebox{0}{\includegraphics{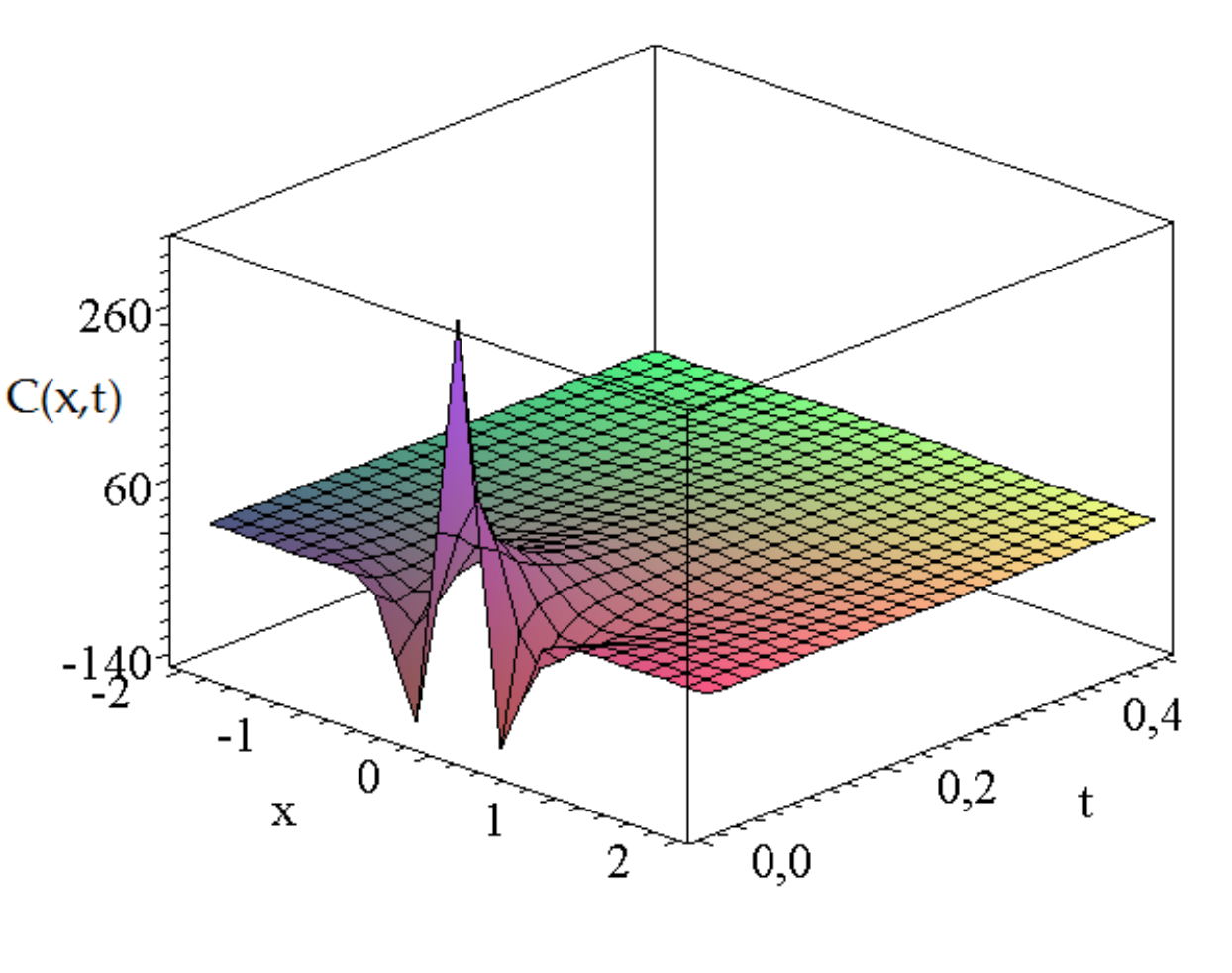}}}
\caption{The solution $C(x,t)$ for $\alpha = 3/2$ in the form of Eq.  (\ref{haromketted}). }
\label{kettes}      
\end{figure}

For the {\it even solutions}, we denote by $g(\eta)$ the following function 
\begin{equation} 
f(\eta) = \eta  \cdot e^{-\frac{ \eta^2}{4D} }  g(\eta),  
\label{even}
\end{equation} 
Inserting this equation into Eq. (\ref{ode1}), we have 
\begin{equation} 
\eta g'' +2 g' - \frac{\eta^2}{2D} g' + (\alpha-1) \frac{\eta}{D} g = 0.
\label{eq-geta}
\end{equation}  
In concordance with Eq. (\ref{f_eta2}), we get the general solution 
\begin{equation} 
g(\eta) = \left(  c_1 M\left[1-\alpha  , \frac{3}{2} , \frac{ \eta^2}{4D} \right]  + c_2 U\left[1 -\alpha , \frac{3}{2} , \frac{ \eta^2}{4D} \right]    \right). 
\end{equation}  
At this point we make the conjecture from the forms of $U$ and $M$, that if we had the classical spatially even solution for $\alpha=1/2$, than the next 
spatially even solution is for $\alpha = 3/2$, with the form of $g$   
\begin{equation} 
g(\eta) = K_0 \frac{1}{\eta} + K_1 \eta,
\label{geta1}
\end{equation} 
where $K_0$ and $K_1$ are arbitrary constants, which should be determined later. We insert this form of $g$ in (\ref{eq-geta}) we find, that the form (\ref{geta1}) fulfill the equation (\ref{eq-geta}), if  
\begin{equation}
K_1 = - \frac{1}{2D} K_0. 
\end{equation} 
We obtain the same result if we insert the form 
\begin{equation} 
f(\eta) = \eta  \cdot e^{-\frac{ \eta^2}{4D} } 
\left( K_0 \frac{1}{\eta} + K_1 \eta \right), 
\end{equation}
directly into the equation (\ref{ode1}). 
By this, for $\alpha = 3/2$, we get for the function $f$ 
\begin{equation} 
f(\eta) = K_0 \cdot \eta  \cdot e^{-\frac{ \eta^2}{4D} } 
\left(\frac{1}{\eta} - \frac{1}{2D} \eta \right) 
=   K_0  \cdot e^{-\frac{ \eta^2}{4D} } 
\left( 1 - \frac{1}{2D} \eta^2 \right).  
\end{equation}  
Substituting this form into (\ref{ansatz}) one gets 
\begin{equation} 
C(x,t) = K_0 \frac{1}{t^{\frac{3}{2}}} 
e^{-\frac{ x^2}{4 D t} } 
\left( 1 - \frac{1}{2D} \frac{x^2}{t} \right). 
\label{haromketted}
\end{equation}
This result is visualized on Figure  (\ref{kettes}). \\

\begin{figure}  
\scalebox{0.4}{
\rotatebox{0}{\includegraphics{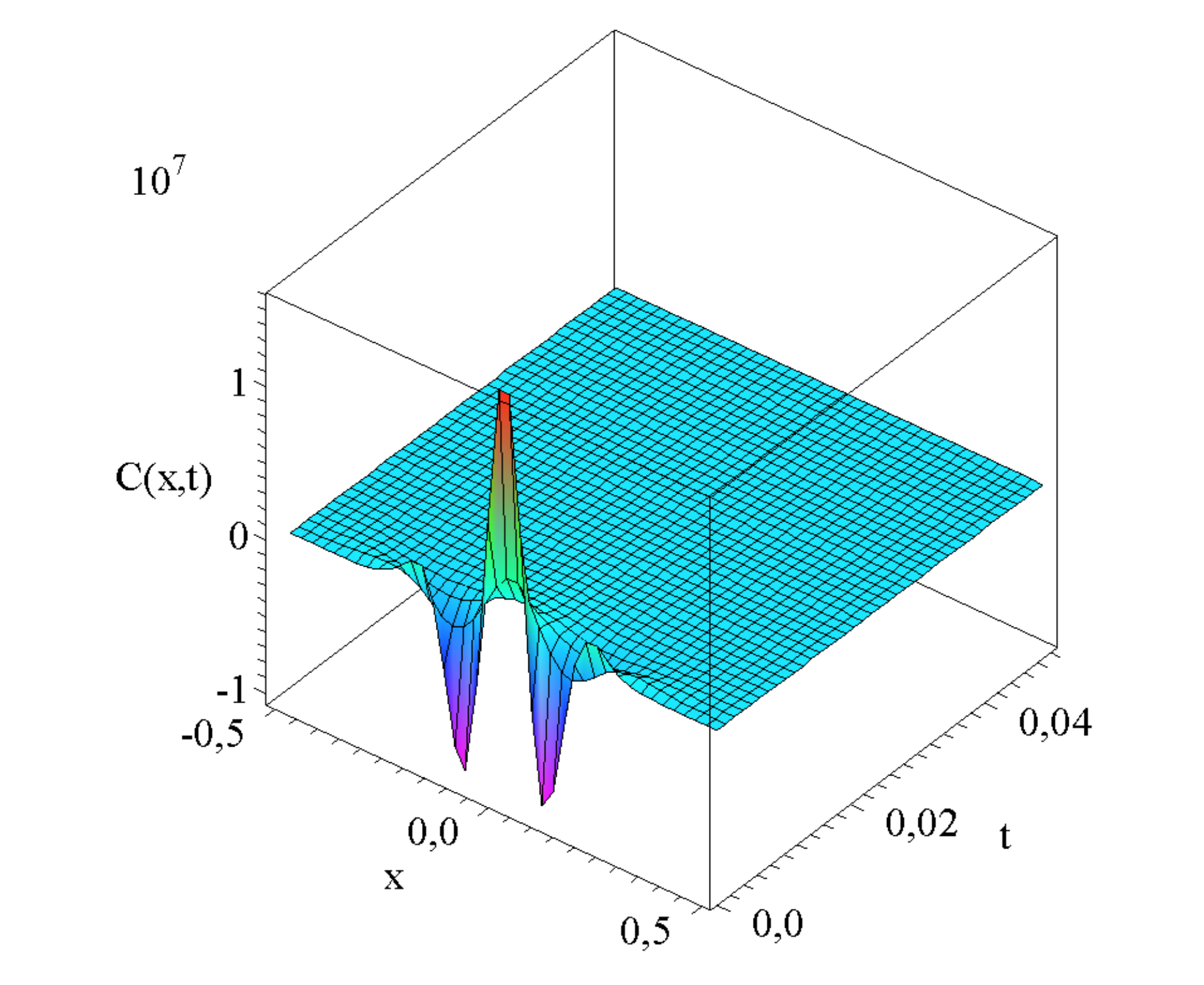}}}
\caption{The solution $C(x,t)$ for $\alpha = 5/2$ in the form of Eq.  (\ref{otketted}), respectively.}
\label{harmas}      
\end{figure}

If we follow the case $\alpha = 5/2 = 2.5$, then the following 
form for the function $g(\eta)$ can be considered
\begin{equation} 
g(\eta) = K_0 \cdot \frac{1}{\eta} + K_1 \cdot \eta + K_2 \cdot \eta^3.
\label{series}
\end{equation}
If we insert this form in the equation (\ref{eq-geta}) 
the following relations for the constants $K_0$, $K_1$ and $K_2$ 
can be derived 
\begin{equation} 
K_1 = \frac{K_0}{D}, 
\end{equation}
and
\begin{equation} 
K_2 = - \frac{K_1}{12 D} = \frac{K_0}{12 D^2}.
\end{equation} 
By this, we get for the $g(\eta)$ 
\begin{equation} 
g(\eta) = K_0 \left( \frac{1}{\eta} - \frac{1}{D} \eta + 
                     \frac{1}{12 D^2} \eta^3  \right). 
\end{equation}
Correspondingly the final form for $f(\eta)$ for $\alpha = 2.5$ is 
\begin{equation}
f(\eta) = K_0 \cdot e^{-\frac{ \eta^2}{4D} }
 \left( 1 - \frac{1}{D} \eta^2 + 
                     \frac{1}{12 D^2} \eta^4  \right).
\end{equation}
Inserting this form into (\ref{ansatz}) one gets 
\begin{equation} 
C(x,t) = K_0 \frac{1}{t^{\frac{5}{2}}} 
e^{-\frac{ x^2}{4 D t} } 
\left( 1 - \frac{1}{D} \frac{x^2}{t} + 
                     \frac{1}{12 D^2} \frac{x^4}{t^2}  \right). 
\label{otketted}
\end{equation}
This result can be seen on Figure (\ref{harmas}).  \\ 

\begin{figure}  
\scalebox{1.0}{
\rotatebox{0}{\includegraphics{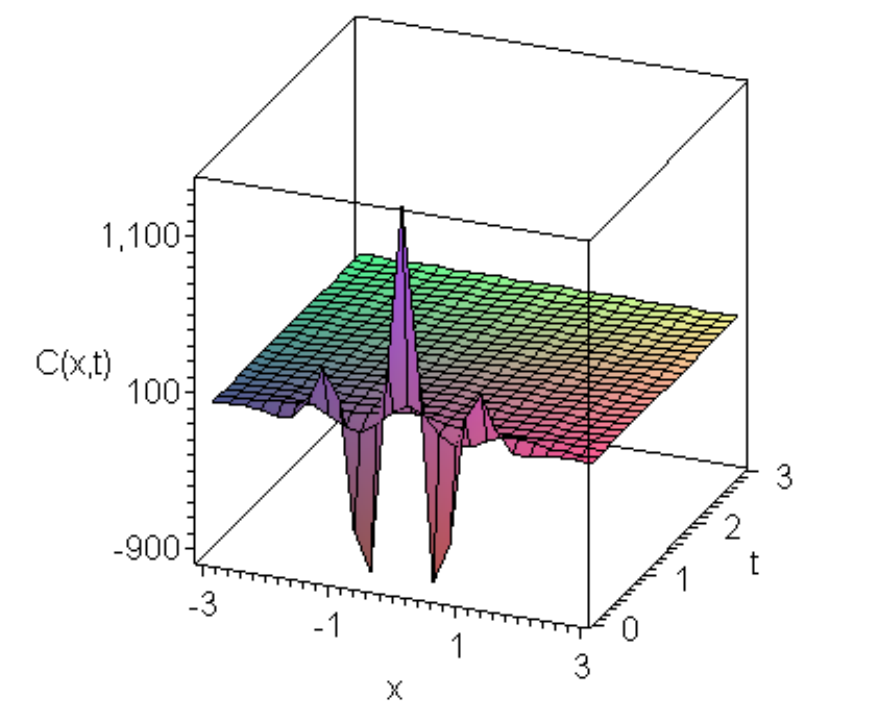}}}
\caption{ The function $C(x,t)$ for the value 
$\alpha=7/2$.    }
\label{fig-hetketted}      
\end{figure}

If we follow the case $\alpha = 7/2 = 3.5$, then the following 
form for the function $g(\eta)$ can be considered: 
\begin{equation} 
g(\eta) = K_0 \cdot \frac{1}{\eta} + K_1 \cdot \eta + K_2 \cdot \eta^3 + K_3  \cdot \eta^5.
\label{series7-2}
\end{equation}
If we replace this form into the equation (\ref{eq-geta}) 
the next relations among the constants $K_0$, $K_1$, $K_2$ and $K_3$ can be derived: 
\begin{equation} 
K_1 = - \frac{3}{2} \frac{K_0}{D} ,  
\end{equation}
for the next coefficient 
\begin{equation} 
K_2 = -\frac{K_1}{6 D} = \frac{K_0}{4 D^2}. 
\end{equation}
Finally for the third coefficient one gets 
\begin{equation} 
K_3 = - \frac{K_2}{30 D} = - \frac{K_0}{120 D^3}.
\end{equation}
Inserting these coefficients into the formula 
(\ref{series7-2}), one obtains the following expression
\begin{equation} 
g(\eta) = K_0 \bigg( \frac{1}{\eta} - \frac{3}{2D} \cdot \eta + \frac{1}{4 D^2} \cdot \eta^3 
- \frac{1}{120 D^3}  \cdot \eta^5 \bigg).
\label{series7-2-final}
\end{equation}
This form of $g$ yields, by eq. (\ref{even}), for the function $f$ 
\begin{equation} 
f(\eta) = K_0 \cdot e^{-\frac{ \eta^2}{4D} }
 \left( 1 - \frac{3}{2D} \eta^2 
 +\frac{1}{4 D^2} \eta^4 - \frac{1}{120 D^3} \eta^6  \right).
\end{equation}
Inserting this form into (\ref{ansatz}) one gets 
\begin{equation} 
C(x,t) = K_0 \frac{1}{t^{\frac{7}{2}}} 
e^{-\frac{ x^2}{4 D t} } 
\left( 1 - \frac{3}{2D} \frac{x^2}{t} + 
\frac{1}{4 D^2} \frac{x^4}{t^2}   
- \frac{1}{120 D^3} \frac{x^6}{t^3} \right). 
\label{hetketted}
\end{equation}
This result is clearly visualized on Figure \ref{fig-hetketted}.  

It is evident that including higher terms in the finite series of Eq. (\ref{series7-2}) 
the solutions for $\alpha = 9/2, 11/2, $ etc. can be evaluated in a direct way.  

\begin{figure}  
\scalebox{0.55}{
\rotatebox{0}{\includegraphics{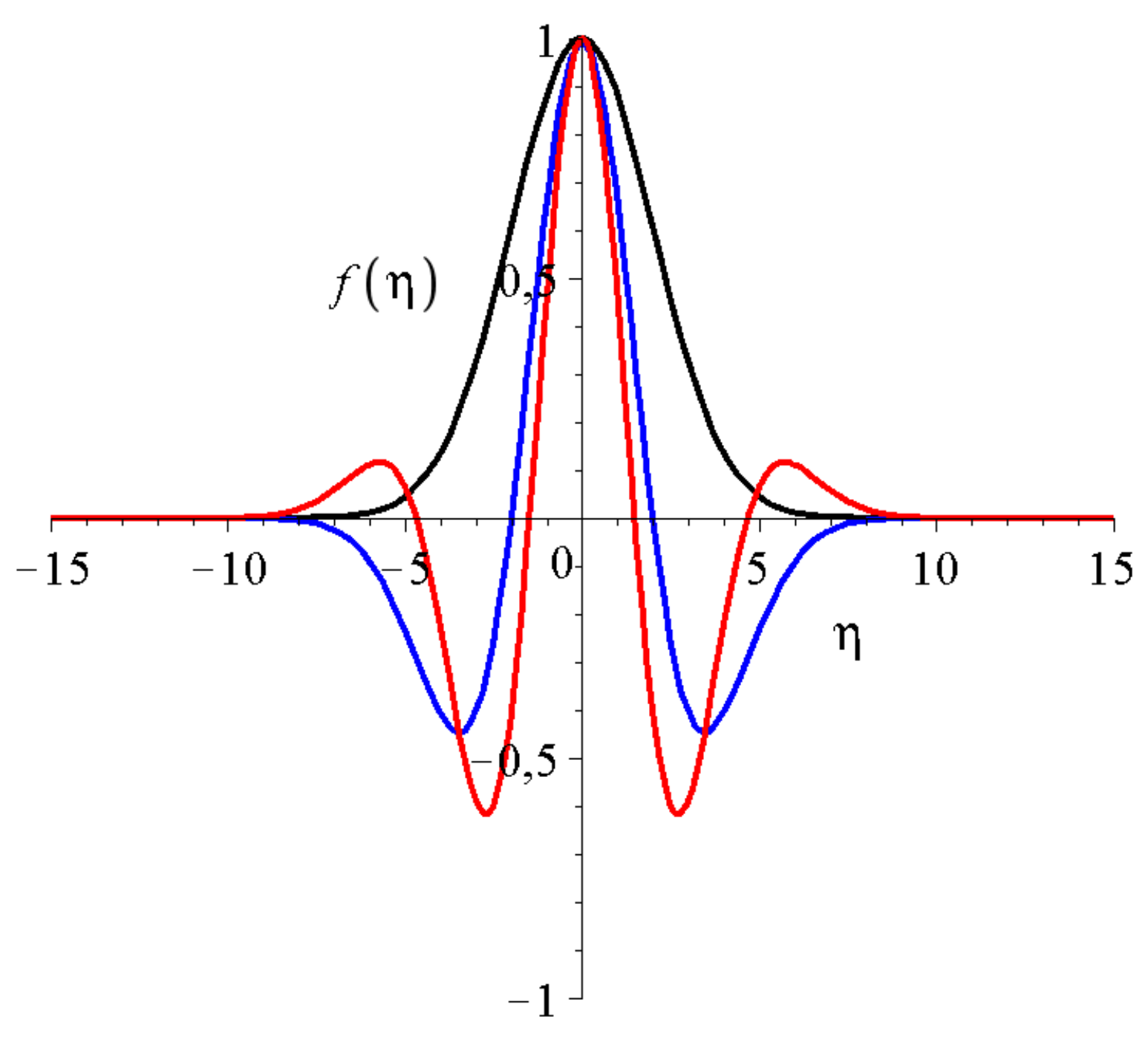}}}
\caption{Even shape functions  $f(\eta)$ of Eq. (\ref{even}) for three different 
self-similar $\alpha$ exponents.  The black, blue and red curves are for $\alpha = 1/2, 3/2$ and $5/2$ numerical values, with the same diffusion constant ($D$ = 2), 
 respectively. Note, shape functions with larger $\alpha$s have more zero transitions. We will show that 
 for $\alpha > 0$ integer values, the integral of the shape functions give zero on the whole and the half-axis as well. }
\label{negyes}      
\end{figure}

For completeness we present the shape functions $f(\eta)$s on Figure \ref{negyes}.
Note, that solutions with higher $\alpha$ values have more oscillations and quicker decay. The same features appear for odd solutions as well.  

As we can see, at this point, the solutions fulfills the 
boundary condition $C() \rightarrow 0$ if $x \rightarrow \pm \infty$, 
for positive $\alpha$ values.

The general initial value problem can be solved with the 
usage of the Green's functions formalism. According to the standard theory of the Green's functions the solution of the 
diffusion equation (\ref{pde}) can be obtained via the next convolution integral: 
\eq
C(x,t) = \frac{1}{2\sqrt{\pi t}} \int_{-\infty}^{+\infty} w(x_0) G(x-x_0)dx_0, 
\eqe
where $w(x_0)$ defines the initial condition of the problem, $C\vert_{t=0} = w(x_0)$ . 
The Green's function for diffusion is well-defined and can be found in many mathematical textbooks eg. \cite{green1,green2,greiner,bronstein}, 
\eq
G(x-x_0) = exp\left[- \frac{(x-x_0)^2}{4t D}\right]. 
\label{gau_green}
\eqe

On the other side, the Gaussian function is a fundamental solution of diffusion.  \\ 

  We will see in the following that for some special forms of the initial conditions, like polynomials, 
Gaussian, Sinus or Cosines the convolution integral can be done analytically. 

In the following we evaluate the convolution integral for $\alpha = 1/2$. 

\begin{figure}  
\scalebox{0.8}{
\includegraphics{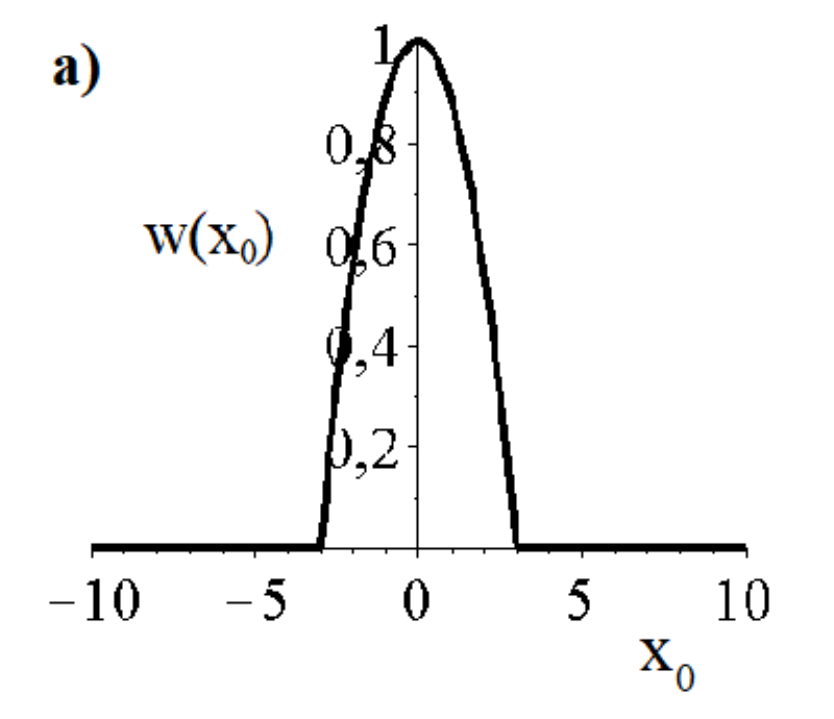}}
\hspace{1mm}
\scalebox{0.8}{
\rotatebox{0}{\includegraphics{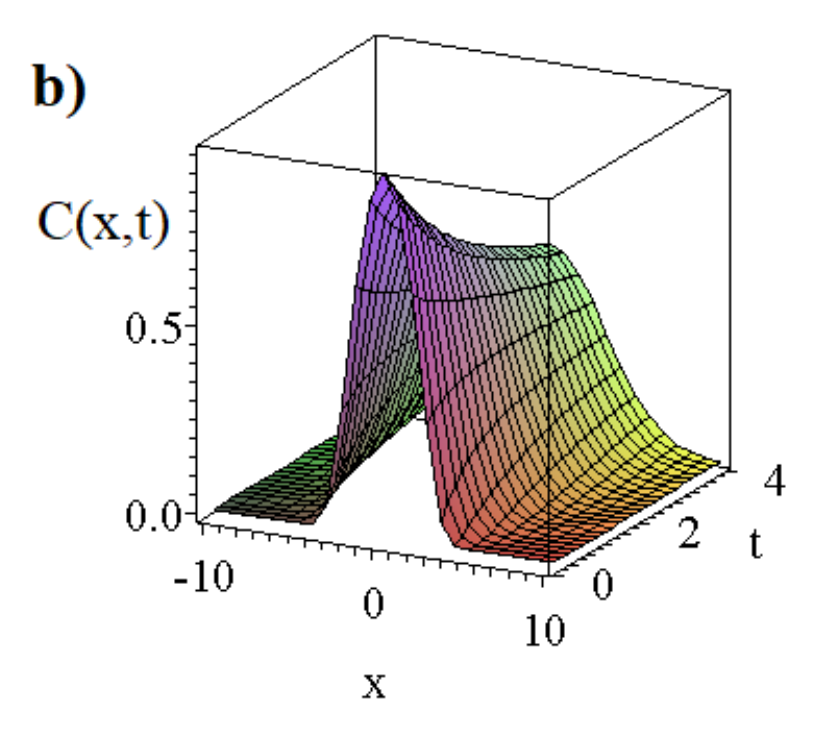}}}
\caption{a) The initial condition (\ref{init}) b) The convolution integral for $\alpha=1/2$ 
of Eq. (\ref{convol-0.5}).}
\label{GreenFv-0.5}      
\end{figure}

As an example for the initial condition problem we may consider the following smooth function with a compact support: 
\eq
w(x_0)= \frac{\text{Heaviside}(3-x_0)\cdot \text{Heaviside}(3+x_0) \cdot (9-x_0^2)}{9}.
\label{init}
\eqe
This initial condition is a typical initial distribution for diffusion, and one can see on 
Figure \ref{GreenFv-0.5}a).

The convolution integral for $\alpha =1/2$: 
\begin{equation} 
C(x,t) = \frac{1}{2 \sqrt{\pi t}} 
\int_{-\infty}^{+\infty} 
\frac{\text{Heaviside}(3-x_0)\cdot \text{Heaviside}(3+x_0) \cdot (9-x_0^2)}{9} \cdot 
e^\frac{(x-x_0)^2}{4Dt} dx_0.  
\label{convol-0.5}
\end{equation}
The result of this evaluation is 
\begin{eqnarray} 
&C(x,t) = \frac{1}{2 \sqrt{\pi t} } 
\bigg[ 
\sqrt{\pi t} \ \text{erf}
\left( \frac{3+x}{2 \sqrt{ t}} \right) 
+\frac{2}{9} x t \ e^{-\frac{6x+x^2+9}{4t} } 
+ \frac{2}{3} t \ e^{-\frac{6x+x^2+9}{4t} }
-\frac{2}{9} t^{\frac{3}{2}} \sqrt{\pi} \ \text{erf} \left(  \frac{3+x}{2 \sqrt{t}} \right) \nonumber \\
&- \frac{1}{9} x^2 \sqrt{\pi t} \ \text{erf} 
\left( \frac{3+x}{2 \sqrt{t}} \right) 
- \sqrt{\pi t} \ \text{erf}
\left( \frac{x-3}{2\sqrt{t}} \right) 
-\frac{2}{9} x t \ e^{-\frac{-6x+x^2+9}{4t} } 
+ \frac{2}{3} t \ e^{-\frac{-6x+x^2+9}{4t} }
\nonumber \\ 
&+\frac{2}{9} t^{\frac{3}{2}} \sqrt{\pi} \ \text{erf}  
\left( \frac{x-3}{2\sqrt{t}} \right) 
+ \frac{1}{9} x^2 \sqrt{\pi t} \ \text{erf} 
\left( \frac{x-3}{2\sqrt{t}} \right) 
\bigg],
\end{eqnarray}
which is presented on Figure \ref{GreenFv-0.5}b).

\section{The properties of the shape functions and solutions} 

In the following we study the some properties of the shape functions 
$f(\eta)$ and of the complete solutions $C(x,t)$. 
First we consider the $L^1$ integral norms. 

For the case $\alpha = 1/2$ the form of 
\begin{equation} 
\int_{-\infty}^{\infty} f(\eta) d\eta 
=  \int_{-\infty}^{\infty} f_0 e^{-\frac{\eta^2}{4D}}  d\eta 
=   f_0 \, 2  \sqrt{\pi D}. 
\end{equation}
The constant $f_0$ is chosen, depending on the problem. If $C$ stands for the density which diffuses, 
$f_0$ in the above integral is related to the total mass of the system. 

Correspondingly 
\begin{equation} 
\int_{-\infty}^{\infty} C(x,t) dx 
=  \int_{-\infty}^{\infty} f_0 
\frac{1}{\sqrt{t}} e^{-\frac{x^2}{4Dt}}  dx 
=   f_0 \, 2  \sqrt{\pi D}. 
\end{equation}

For the case $\alpha=3/2$ 
\begin{equation} 
\int_{-\infty}^{\infty} f(\eta) d\eta = 
\int_{-\infty}^{\infty} 
K_0  \cdot e^{-\frac{ \eta^2}{4D} } 
\left( 1 - \frac{1}{2D} \eta^2 \right) d\eta = 0 .
\end{equation}
It is interesting to see that the integral of  first even shape function beyond Gaussian is zero. 
An even more remarkable feature is however, that 
\begin{equation} 
\int_{-\infty}^{0} f(\eta) d\eta = \int_{0}^{\infty} f(\eta) d\eta = 0.\end{equation}
So the oscillations, the positions of the zero transitions divide the 
function in such a way that the integral not only the whole real axis $(-\infty ... \infty )$ but on the half axis $(0..\infty) $ or $(-\infty.. 0$) gives zero as well. 

Evaluating the same type of integrals for the   
corresponding solution $C(x,t)$ we have 
\begin{equation} 
\int_{-\infty}^{\infty} C(x,t) dx = \int_{0}^{\infty} C(x,t)dx =  \int_{-\infty}^{0} C(x,t)dx = \nonumber \\
\int_{-\infty}^{\infty}  
K_0  \cdot \frac{1}{t^{3/2}} e^{-\frac{ x^2}{4Dt} } 
\left( 1 - \frac{1}{2D} \frac{x^2}{t} \right) dx = 0, 
\end{equation}
at any time point, (and for any diffusion coefficient D). 

\begin{figure}  
\scalebox{0.8}{
\rotatebox{0}{\includegraphics{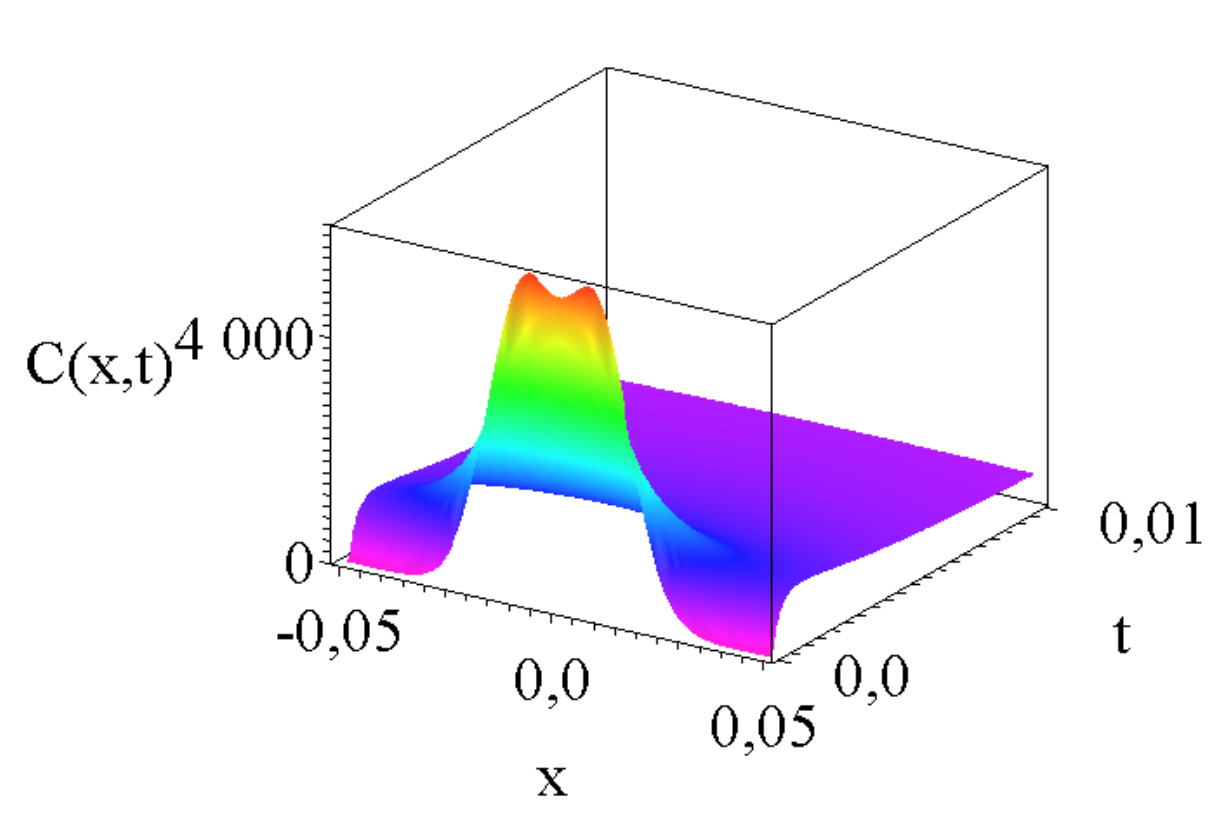}}}
\caption{The function $C(x,t)$, solution of Eq. (\ref{lineq})}
\label{kilenc}      
\end{figure}

The same property is true for all possible higher harmonic solutions if $\alpha$ is positive half-integer number  
$ \alpha =  (2n+1)/2$ when $ (n \> \epsilon  \>  \mathbb{N}  $). 
This property has far-reaching consequences. 
The linearity of the regular diffusion equation and 
this additional property of this even series of solutions makes it possible to perturb the usual Gaussian in such a way, that the 
total number of particles are conserved during the diffusion process, 
however the initial distribution can be changed significantly. 
One can see from the final form of the solutions 
$ C(x,t)_{\alpha} \sim \frac{1}{t^{\alpha}}$ that the decay of these perturbations are however short-lived because they have a quicker decay than the standard Gaussian solutions. 
For completeness we present a $C(x,t)$ solutions which is a linear combination 
of the first two even solutions  $\alpha = 1/2, 3/2$  in the form of 
\begin{equation} 
C(x,t) = \frac{60}{t^{\frac{1}{2}}}e^{-\frac{-x^2}{4t}} - \frac{0.001}{t^{\frac{3}{2}}}e^{-\frac{-x^2}{4t}}\left( 1 - \frac{x^2}{2t}\right),
\label{lineq}
 \end{equation} 
on Fig. (\ref{kilenc}). 
Note, that coefficients with different orders of magnitude had to be applied to reach a visible effect when the sum of two functions have to be visualised with different power-law decay. 

As a second property we investigate the cosine Fourier transform of the shape functions:
\eq
C_{\alpha}(k) = \int_{-\infty}^{\infty} Cos(k\cdot \eta) f_{\alpha}(\eta) d\eta. 
\eqe
In can be shown with direct integration, that the Fourier 
transform is  
\eq
C_{\alpha = \frac{2N+1}{2}}(k) \propto  l\cdot \sqrt{\pi} \cdot  \frac{  k^{2N}  \cdot D^{N}  \cdot e^{-k^2 D}}{\sqrt{\frac{1}{D}}},  
\eqe
for all $N \> \epsilon \> \mathbb{N}  \symbol{92}0 $  positive integer and l is a real constant. 
This means, that qualitatively the spectra for all positive half integer $\alpha$ are similar. 
They start from zero, have a global positive maximum and and a quick decay to zero. 
It is generally known from spectral analysis that pulses of finite length have 
band spectra which have a minimal a maximal and a central frequency.  \\ 
In Appendix A the corresponding normalization 
coefficients are given for the odd functions as well.

\section{The diffusion equation with constant source}

At this point we try to find solutions of the 
diffusion equation, mainly with the self similar Ansatz, where on the right hand side, there is a constant source term. 
\begin{eqnarray}
\frac{\partial C(x,t)}{\partial t  } = D \frac{\partial^2 C(x,t)}{\partial x^2 } + n, 
\label{pde-source} 
\end{eqnarray}
For this equation one also apply the self-similar transformation (\ref{ansatz}), and we get a 
modified equation relative to the homogeneous one 
\eq
-\alpha t^{-\alpha -1} f(\eta)  - \beta   t^{-\alpha -1} \eta  
\frac{d f(\eta)}{d\eta}  =  D  t^{-\alpha -2\beta}  \frac{d^2f(\eta)}{ d\eta^2} + n .   
\label{ode-source}
\eqe
The free term on the r.h.s. has no explicit time decay, consequently we expect the same from the other terms, which means 
\begin{eqnarray}
-\alpha - 1 & = & 0  \\ 
-\alpha -2\beta & = & 0 .
\end{eqnarray}
The two equations have to be fulfilled simultaneously. Solving these equations, we get the following values for $\alpha$ and  $\beta$: 
\begin{equation} 
\alpha = -1  \,\, \textrm{and} \,\, \beta=\frac{1}{2}
\end{equation}

Inserting these values to the equation 
(\ref{ode-source}), we get the following ODE  
\eq
f(\eta)  - \frac{1}{2}    \eta  
\frac{d f(\eta)}{d\eta}  =  D  \frac{d^2f(\eta)}{ d\eta^2} + n .   
\label{ode-source2}
\eqe
We emphasize, that we arrived to this equation by a 
self-similar transformation.  
At this point we observe, that if we shift the function $f$ by a constant, and introduce the function $h$: 
\begin{equation} 
h(\eta) = f(\eta) - n 
\end{equation} 
we arrive to a slightly modified equation 
\eq
h(\eta)  - \frac{1}{2}    \eta  
\frac{d h(\eta)}{d\eta}  =  D  \frac{d^2 h(\eta)}{ d\eta^2} . 
\label{ode-source2-h}
\eqe
One may observe, that if the transformation 
$\eta \rightarrow -\eta $ and $h(-\eta)=h(\eta)$
is applied, the equation still remains the same, 
consequently we expect at least one even solution. 

If we look for the even solution by polynomial expansion 
\begin{equation}
h(\eta) = A + B \eta^2 + ...
\end{equation}
then we get by direct substitution 
\begin{equation} 
A = 2 \cdot B \cdot D . 
\end{equation}
This means, that the even solution reads as follows
\eq 
h(\eta) = B ( 2 D +  \eta^2 ) 
\eqe
where $B$ is a constant depending on initial conditions. 

\begin{figure}  
\scalebox{0.8}{
\rotatebox{0}{
\includegraphics{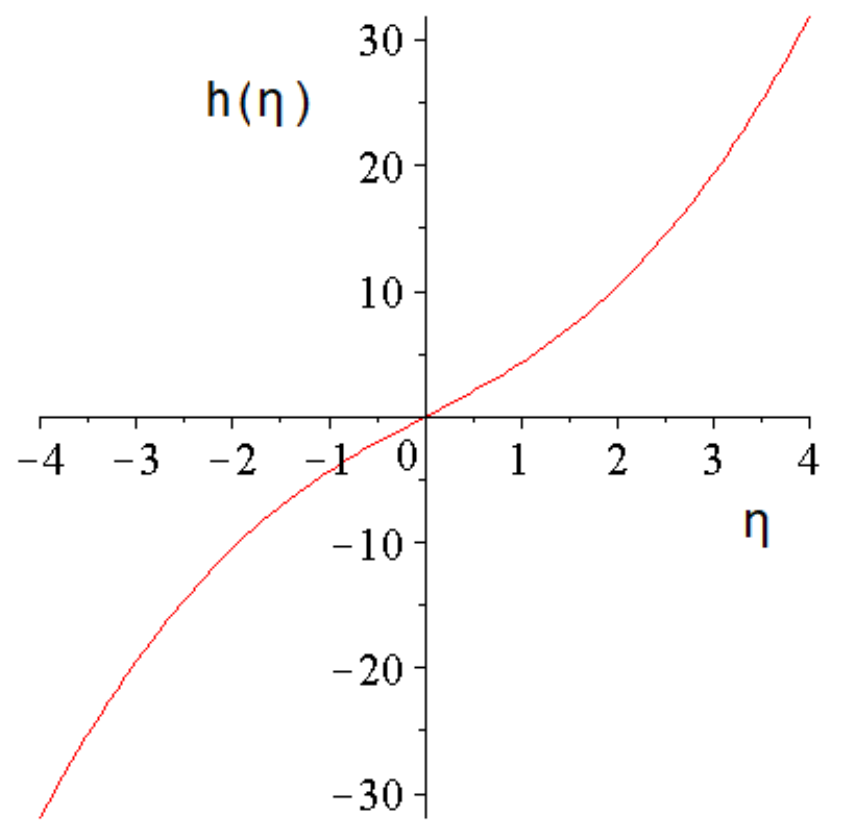}}}
\caption{The shape function $h(\eta)$, described by Eq. (\ref{eq:odd-source}), the odd solution of 
Eq. (\ref{ode-source2-h}).}
\label{fig:odd-sol}      
\end{figure}

Furthermore, we observe, that 
the transformation 
$\eta \rightarrow -\eta $ and $h(-\eta)=-h(\eta)$
also leaves the equation (\ref{ode-source2-h})  unchanged. This means, that it is worthwhile to look for an odd solution, too. 
The odd solution of the equation is 
\begin{equation} 
h(\eta)= 2 D \, \eta \, e^{-\frac{\eta^2}{4 D}}  
+ \sqrt{\pi} \, (2 D^{3/2} +\sqrt{D}\, \eta ^2) \, 
erf \left( \frac{1}{2} \frac{\eta}{\sqrt{D}} \right)
\label{eq:odd-source}
\end{equation}
The form of this odd solution one can see on Figure 
\ref{fig:odd-sol}. 

If $n$ is positive in the equation (\ref{pde-source}), then we can talk about a source in the 
equation, and if $n$ is negative, than we say that there is a sink in the diffusion process. 
The sink can be considered physical by the time 
$C(x,t) \geq 0$. Diffusive systems with sinks have been studied in ref. \cite{ClGa2001}, and water purification by adsorption also means a process with change of concentration is space and decrease in time \cite{RaTo2021}. 

The general solution for the shape function can be obtained from the linear combination of the 
even and odd solutions presented above  
\begin{equation} 
h(\eta) = \kappa_1 \bigg[ 
2 D \, \eta \, e^{-\frac{\eta^2}{4 D}}  
+ \sqrt{\pi} \, (2 D^{3/2} +\sqrt{D}\, \eta ^2) \, 
erf \left( \frac{1}{2} \frac{\eta}{\sqrt{D}} \right)
\bigg] + \kappa_2 [ 2 D +  \eta^2]
\end{equation}
where $\kappa_1$ and $\kappa_2$ are constants depending on 
the initial or boundary conditions of the problem. 

Inserting this shape function to the general solution (\ref{ansatz}), we get for the final form of $C(x,t)$ in the presence of a constant source
\begin{equation} 
C(x,t) = t \cdot 
\bigg[ \kappa_1 \bigg( 
2 D \, \eta \, e^{-\frac{x^2}{4 D t}}  
+ \sqrt{\pi} \, \left( 2 D^{3/2} +\sqrt{D}\,  
\frac{x^2}{t} \right) \, 
erf \left( \frac{1}{2} \frac{x}{\sqrt{D t}} 
\right)
\bigg) + \kappa_2 
\left( 2 D +  \frac{x^2}{t} \right) +n \bigg]
\label{eq:final-source}
\end{equation}

\begin{figure} 
\scalebox{0.8}{ 
\includegraphics{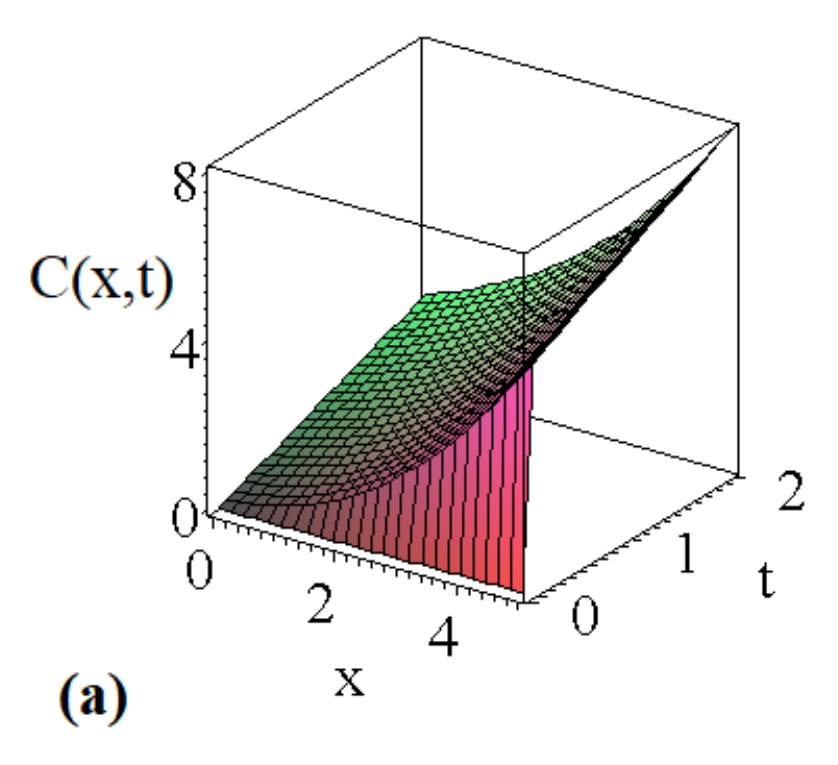}}
\hspace{0.1cm}
\scalebox{0.8}{
\rotatebox{0}{
\includegraphics{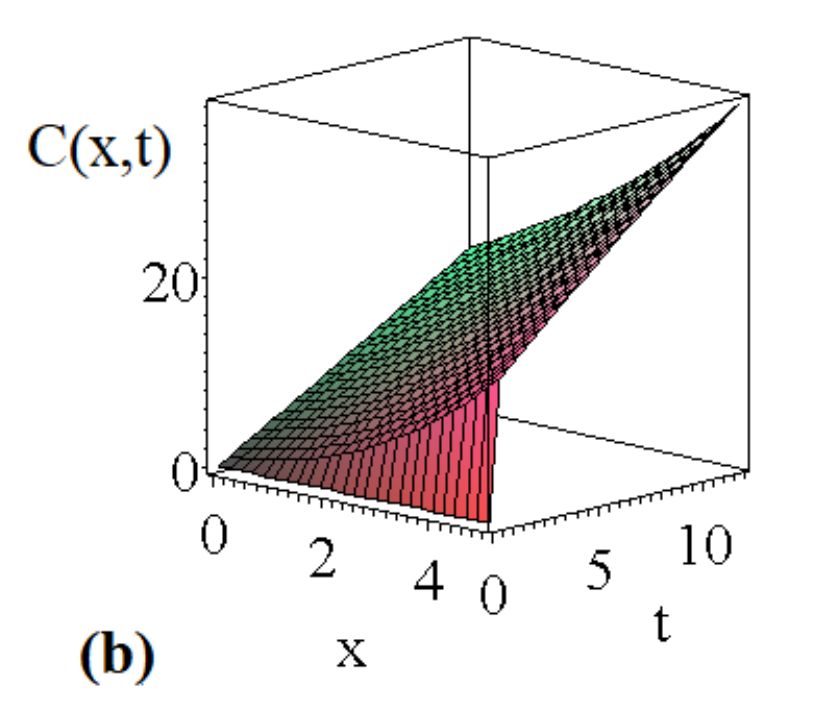}}}
\caption{The shape function $C(x,t)$, solution of Eq. (\ref{eq:final-source}), for $D=1$ and $n=1$, in case a)  $\kappa_1=0.1$ $\kappa_2=0.03$ and b) $\kappa_1=0.2$ $\kappa_2=0.2$. } 
\label{fig:short-transient}      
\end{figure}

For relatively shorter times, the general solution has interesting features depending on the weight of even or the odd part of the solution, as one can see on figure \ref{fig:short-transient}.  

The long time behavior is dominated by the constant of the even solution and the source term. Correspondingly for sufficiently long times the relation  
$C(x,t) \sim (2\kappa_2 \, D + n) \cdot t $ characterizes the dynamics, as one can see on 
Figure \ref{fig:short-transient}b).

 \section{Summary and Outlook }
Applying the well-known self-similar Ansatz - together with and additional 
change of variables - we derived symmetric solutions for the one dimensional diffusion equations. Using the Fourier series analogy we might say that these solutions may be considered as possible higher harmonics of the fundamental Gaussian solution. 
As unusual properties we found that the integral of these solutions give zero on both 
the half and the whole real axis as well. Thanks to the linearity of the diffusion equation these kind of functions can be added to the 
particle (or energy) conserving fundamental Gaussian solution therefore 
new kind of particle diffusion processes can be described. Due to the higher $\alpha$
self-similar exponents these kind of solutions give relevant contributions only at smaller time coordinates, because the corresponding solutions  decay quicker than the usual Gaussian solution.  \\
These kind of solutions can be also evaluated for two or three dimensional, cylindrical or spherical symmetric systems as well. 
Work is in progress to apply this kind of analysis to more sophisticated diffusion systems as well. 
We hope that our new solutions have far reaching consequences and they  
will be successfully applied in other scientific disciplines like quantum mechanics, quantum field theory in physics, in probability theory or in financial mathematics in the near future.

\section{Acknowledgments} One of us (I.F. Barna) was supported by
the NKFIH, the Hungarian National Research
Development and Innovation Office.\\ 
- The authors declare no conflict of interest. \\
- Both authors contributed equally to every detail of the study. \\
-There was no extra external founding.

 \section{Appendix}
For completeness and for direct comparison we show the first five odd shape functions $f(\eta)$ and the corresponding solutions $C(x,t)$:  

\begin{eqnarray}
f(\eta) &=&      erf \left({\frac{\eta}{2\sqrt{ D}}} \right), \nonumber \\ 
f(\eta) &=& \kappa_0 \cdot \eta \cdot  e^{-\frac{\eta^2}{4D}},  \nonumber \\ 
f(\eta) &=&  \kappa_0 \cdot \eta  \cdot   e^{-\frac{\eta^2}{4D}}  \cdot \left( 1 -  \frac{1}{6D} \eta^2   \right), \nonumber  \\    
f(\eta) &=& \kappa_0 \cdot \eta \cdot e^{-\frac{\eta^2}{4D}} \cdot \left( 1 - \frac{1}{3D} \eta^2 
                         + \frac{1}{60} \frac{1}{D^2}  \eta^4 \right), \nonumber \\  
f(\eta) &=& \kappa_0 \cdot \eta \cdot e^{-\frac{\eta^2}{4D}} \cdot \left( 1 - \frac{1}{2D} \eta^2 
                         + \frac{1}{20} \frac{1}{D^2}  \eta^4 - \frac{1}{840} \frac{1}{D^3}  \eta^6   \right),
\label{fs}
\end{eqnarray}
    for $\alpha = 0,1,2,3,4..\mathbb{N}  $. 
    The first case with the change of variable 
    $x/\sqrt{t}$ with no $\alpha$ (or implicitly  $\alpha=0$) dates back to 
    Boltzmann \cite{Boltzmann1894}, as it is also  mentioned by \cite{Lonngren1977} and 
    \cite{MeSt2009}. 
    
    All integrals of the functions from (\ref{fs}) on 
    the whole real axis give zero: 
    \eq
  \int_{-\infty}^{\infty} f_{\alpha}(\eta) d\eta = 0, 
  \eqe
 however on the half-axis: 
 \eq \int_{0}^{\infty} f_{\alpha = 0}(\eta) d\eta = \infty, 
 \eqe 
 and for additional 
 non-zero integer $\alpha$s we get: 
 \eq \int_{0}^{\infty} f_{\alpha}(\eta) d\eta =  \frac{D}{\alpha - 1/2}. 
 \eqe
 Integrals on the opposite half-axis $(-\infty..0]$ have the same value with a negative 
 sign, respectively.  
The forms for odd  $C(x,t)$s are the following:
 \begin{eqnarray}
C(x,t) &=&      erf \left({\frac{x}{2\sqrt{ Dt}}} \right), \nonumber \\ 
C(x,t) &=&     \left( \frac{\kappa_1 x}{t^{\frac{3}{2}}}  \right)   e^{-\frac{x^2}{4Dt}}, \nonumber \\ 
C(x,t )&=& \left( \frac{\kappa_1 x}{t^{\frac{5}{2}}}  \right)    e^{-\frac{x^2}{4Dt}}  
 \left( 1 -  \frac{x^2}{6Dt}  \right), \nonumber  \\  
C(x,t) &=&     \left( \frac{\kappa_1 x}{t^{\frac{7}{2}}}  \right)  e^{-\frac{x^2}{4Dt}} \left( 1 - \frac{x^2}{3Dt}  
                         + \frac{x^4}{60(Dt)^2}    \right), \nonumber \\  
C(x,t) &=&    \left( \frac{\kappa_1 x}{t^{\frac{9}{2}}}  \right)    e^{-\frac{x^2}{4Dt}}  \left( 1 - \frac{x^2}{2Dt}  
                         + \frac{x^4}{20(Dt)^2}  - \frac{x^6}{840(Dt)^3}    \right),
\label{cs}
 \end{eqnarray}
The space integrals of $\int_{-\infty}^{\infty} C_{\alpha }(x,t) dx = 0 $ for all positive integer $\alpha$s. 
On the positive half-axis for $\alpha = 0$ the integral of the error function in infinite, for positive $\alpha$a it is:  
\begin{equation}
\int_{-\infty}^{\infty} C_{\alpha}(x,t)  = \frac{D t^{\frac{1}{2} -\alpha}}{\alpha -\frac{1}{2}}  . 
\end{equation} 
Which are well defined values for finite,  $D$, $t$ and $\alpha$. 
On the $(-\infty..0]$ half axis the sign is opposite. 
 Additional detailed analysis of the odd functions were presented in our former study \cite{imre1}.



\begin{thebibliography}{99}

 \bibitem{crank} J.  Crank, {\it{The Mathematics of Diffusion}}, Oxford, Clarendon Press, 1956.
\bibitem{ghez}  R. Ghez, {\it{Diffusion Phenomena}}, Dover Publication 2001. 
\bibitem{ben} T.D. Bennett, {\it{ Transport by Advection and Diffusion: Momentum, Heat and Mass Transfer}},  John
Wiley \& Sons, 2013.
\bibitem{lien} J.H. Lienhard IV and J.H. Lienhard V, {\it{A heat transfer textbook}}, fourth edition. Cambridge, Massachusetts, USA: Phlogiston Press, 2017.
\bibitem{neu} J. Newman and V. Battaglia, {\it{ The Newman Lectures on Transport Phenomena}},  Jenny Stanford
Publishing, 2021. 
\bibitem{KPZ} M. Kardar, G. Parisi, Y.-C. Zhang, 
Phys. Rev. Lett. {\bf{56}}, 889 (1986).   
\bibitem{BaMa-KPZ1} 
I.F. Barna, G. Bogn\'ar. M. Guedda, K. Hricz\'o, and L. M\'aty\'as, 
Mathematical Modelling and Analysis {\bf{25}}, 241 (2020). 
\bibitem{BaSt1995} 
A.L.Barab\'asi, and E. Stanley, Fractal Concepts in Surface Growth, Cambridge University Press, Cambridge, 1995.  
\bibitem{MaGa2005} 
L. M\'aty\'as and P. Gaspard, 
Physical Review E {\bf{71}}, 036147  (2005) . 
\bibitem{laci}  L. M\'aty\'as and I.F. Barna, 
Romanian Journal of Physics  {\bf{67}},  101 (2022).  
\bibitem{imre1}  I.F. Barna and L. M\'aty\'as, 
Mathematics {\bf{10}}, 3281 (2022).
\bibitem{Cannon1984} 
J.R. Cannon, The one-dimensional heat equation, Addison-Wesley Publishing, Reading, Massachusetts, 1984.   
\bibitem{CoBe2011} 
K.D. Cole, J.V. Beck, A. Haji-Sheikh, B. Litkouhi, Heat Conduction using Green's functions, Series in Computational and Physical Processes in Mechanics and Thermal Sciences (2nd ed), Boca Raton, FL, CRC Press, 2011. 
\bibitem{MaTeVo2000} 
L. M\'aty\'as, T. T\'el and J. Vollmer, Physical Review E {\bf 61}, R3295, 2000. 
\bibitem{Thamby2011} 
R.K.M. Thambynayagam, The Diffusion Handbook: Applied Solutions for Engineers, McGraw-Hill, 2011. 
\bibitem{MiAlRi2015} 
G. Micahaud, G. Alecian, G. Richer, {\it Atomic Diffusion in Stars}; Astronomy and Astrophysics Library; Springer: New York, USA, 2013, Volume 70. 

\bibitem{Murray} 
J.D. Murray, {\it{Mathematical Biology II: Spatial Models and Biomedical Applications}} (Third Edition),  New York, Heidelberg, Springer, 2003. 
\bibitem{Al2023} 
J. Alebraheem, Hindawi Journal of Mathematics {\bf 2023}, 
Article ID 4349573, pp.1-13, (2023). 
\bibitem{Perthame2015} 
B. Perthame, {\it{Parabolic Equations in Biology}},  Springer International Publishing, Heidelberg 2015.
\bibitem{SzeNeKe2017}
R. Sz\'ep, E. Mateescu, A.C. Nechifor, \'A Keresztesi, 
Environmental Science and Pollution Research {\bf 24},
27288  (2017). 
\bibitem{GiSe2013} 
D.T. Gillespie, E. Seitaridou, {\em Simple Brownian Diffusion}, Oxford University Press, Oxford, 2013.
\bibitem{Pernyeszi2009} 
K. T\'alos, Cs. P\'ager, Sz. Tonk, C. Majdik, B. Kocsis, 
F. Kil\'ar, T. Pernyeszi, 
Acta Universitatis Sapientiae: Agriculture
and Environment {\bf 1}, 20 (2009). 
\bibitem{NeNe2009} 
G. Nechifor, S.I. Voicu, A.C. Nechifor, S. Garea,  
Desalination {\bf 241}, 342 (2009). 
\bibitem{LvReCh2018}
J. Lv, K. Ren, and Y. Chen, 
Physical Chemistry {\bf 122}(5), 1655, (2018). 

\bibitem{innov1}  T. H\"agerstrand, {\it{Innovation Diffusion as a Spatial Process}}, The University of Chicago Press, 1967.  
\bibitem{innov2} E. M. Rogers, {\it{Diffusion of Innovations}}, The Free Press 1983.
\bibitem{tech} N. Nakicenovic and A. Gri\"ubler  {\it{Diffusion of Technologies and Social Behavior}}, 
Springer 1991. 
\bibitem{diff_gen1} A. Bunde, J. Caro
J. K\"arger and G. Vogl, {\it{Diffusive Spreading
in Nature, Technology and Society }}, Springer 2018.
\bibitem{diff_gen2}   G. Vogel, {\it{Adventure Diffusion}} , Springer, 2019.
\bibitem{Maz2018} T. Mazzoni, {\it{A First Course in Quantitative Finance}}, 
Cambridge University Press, Cambridge, 2018. 
\bibitem{Ede2014} 
E. L\'az\'ar, 
Acta Universitatis Sapientiae, 
Economics and Business {\bf 2}, 75, (2014). 
\bibitem{Reka-Barabasi-2002} 
R. Albert, and Albert-L\'aszl\'o Barab\'asi, Reviews of Modern Physics 
{\bf 74}, 47 (2002).  
\bibitem{van1} P. Rogolino, R. Kov\'acs, P. V\'an and 
V.A. Cimmelli, Continuum Mech. Thermodyn. {\bf{30}}, 1245–1258 (2018). 
\bibitem{bluman} G.W. Bluman  and  J.D. Cole, Journal of Mathematical Mechanics. {\bf{18}}, 1025 (1969). 
\bibitem{sedov} L. Sedov, {\it{Similarity and Dimensional Methods in Mechanics}} CRC Press, 1993.
\bibitem{zeld} Ya. B. Zel'dovich and Yu. P. Raizer {\it{Physics of Shock 
Waves and High Temperature Hydrodynamic Phenomena}} Academic Press, New York, 1966.
\bibitem{barenb} G.I. Baraneblatt, {\it{Similarity, Self-Similarity, and 
Intermediate Asymptotics}} Consultants Bureau, New York 1979. 
\bibitem{imre2}  I.F. Barna and L. M\'aty\'as,  Fluid. Dyn. Res. {\bf{46}}, 055508 (2014). 
\bibitem{BaPoLo2017} 
I.F. Barna, M.A. Pocsai, S. Lokos, L. M\'aty\'as, Chaos, Solitons and Fractals {\bf 103}, 336, (2017). 
\bibitem{BaPoBa2022} 
I. F. Barna, M.A. Pocsai, and G.G. Barnaf\"oldi, 
Mathematics {\bf 10}, 3220, 2022.  
\bibitem{imre3}  I.F. Barna, M.A. Pocsai and L. M\'aty\'as, Journal of  Generalized Lie Theory and Application 
{\bf{11}}, 1000271  (2017).
\bibitem{Csanad2010} 
M. Csan\'ad and M. Vargyas, Europeand Physical Journal A 
{\bf 44}, 473, (2010). 
\bibitem{BaMa2013} 
I.F. Barna, L. M\'aty\'as, Miskolc Mathematical Notes 
{\bf 14}, 785, (2013).
\bibitem{NaSi19} G.Nath and S. Singh, Journal of Astrophysics and Astronomy {\bf 40}, 50  (2019).
\bibitem{Sa20} P.K.\ Sahu, Brazilian Journal of Physics {\bf 50}, 548 (2020).
\bibitem{KaSuZh20} C.\ Kanchana, Y.\ Su and Y.\ Zhao, Communications in Nonlinear Science and Numerical Simulation {\bf 83}, 105129 (2020). 
\bibitem{NIST} F.W.J. Olver, D.W. Lozier, R.F. Boisvert and C.W Clark, {\it{NIST handbook of mathematical functions}}, Cambridge University Press, 2010.
\bibitem{green1} P.K.  Kythe, {\it{ Green’s Functions and Linear Differential Equations}}, Chapman \& Hall/CRC Applied Mathematics and Nonliner Science; CRC Press: Boca Raton, FL, USA, 2011.
\bibitem{green2}
T. Rother, {\it{ Green’s Functions in Classical Physics}}, Lecture Notes in Physics; Springer International Publishing: 
New York, NY, USA, 2017; Volume 938.
\bibitem{bronstein}I.N.  Bronshtein and K.A. Semendyayev, G. Musiol and  H M\"uhlig, {\it{ Handbook of Mathematics}}; Springer, Wiesbaden, Germany 2007.
\bibitem{greiner} W. Greiner and J. Reinhardt, {\it{ Quantum Electrodynamics}}, Springer: Berlin/Heidelberg, Germany, 2009.
\bibitem{ClGa2001} 
I. Claus, P. Gaspard, Physical Review E {\bf 63}, 036227, 2001.
\bibitem{RaTo2021} 
E. R\'ap\'o, Sz. Tonk, Molecules {\bf 26}, 5419, 2021. 

\bibitem{Boltzmann1894} 
L. Boltzmann, Annalen der Physik {\bf 53}, 959, 1894. 
\bibitem{Lonngren1977}
K.E. Lonngren, Proc. Indian Acad. Sci. {\bf 86}(2), 125, (1977).
\bibitem{MeSt2009} 
H. Mehrer and N.A. Stolwijk, Diffusion Fundamentals {\bf 11}, 1, (2009).

\end{thebibliography}
\end{document}